\documentclass[onecolumn, a4paper, 12pt]{article}

\usepackage{geometry} 

\usepackage{graphicx}
\usepackage{gensymb}
\usepackage{natbib}
\bibpunct{(}{)}{,}{n}{}{}
\makeatletter 
\renewcommand\@biblabel[1]{#1.} 
\makeatother %
\usepackage{fixltx2e}
\usepackage{booktabs}
\usepackage{amsmath,amsfonts,amssymb,epsfig,subfigure}
\usepackage{amsmath,amsfonts,amssymb}

\newcommand{\Keywords}[1]{\par\noindent{\small{\em Keywords\/}: #1}}

\begin{document}

\title{Non-invasive evaluation of skin tension lines with elastic waves}

\author{Claire Deroy$^{a}$, Michel Destrade$^{b, c}$Aidan Mc Alinden$^{d}$,\\ Aisling N\'{i} Annaidh$^{c, e \star,}$. \\[24pt]
$^a$Department of Small Animal Surgery, Veterinary Hospital Fr\'egis, \\94110 Arcueil, France\\[12pt]
$^b$School of Mathematics, Statistics and Applied Mathematics, \\NUI Galway, University Road, Galway, Ireland; \\[12pt]
$^c$School of Mechanical \& Materials Engineering,\\ University College Dublin, Belfield, Dublin 4, Ireland; \\[12pt]
$^d$UCD Veterinary Hospital, \\University College Dublin, Belfield, Dublin 4, Ireland\\[12pt]
$^e$UCD Charles Institute of Dermatology\\School of Medicine and Medical Science, \\University College Dublin, Belfield, Dublin 4, Ireland\\}

\date{}
\maketitle
\pagebreak

\begin{abstract}
\noindent
\emph{Background}\\
Since their discovery by Karl Langer in the 19th Century, Skin Tension Lines (STLs) have been used by surgeons to decide the location and orientation of an incision. Although these lines are patient-specific, most surgeons rely on generic maps to determine their orientation.
Beyond the imprecise pinch test, there remains no accepted method for determining STLs \emph{in vivo}.
\\[4pt] 
\emph{Methods}\\
(i) The speed of an elastic motion travelling radially on the skin of canine cadavers was measured with a commercial device called the Re\-vis\-co\-me\-ter\textsuperscript{\textregistered}. 
(ii) Similar to the original experiments conducted by Karl Langer, circular excisions were made on the skin and the geometric changes to the resulting wounds and excised samples were used to determine the orientation of STLs.
\\[4pt] 
\emph{Results}\\
A marked anisotropy in the speed of the elastic wave travelling radially was observed. The orientation of the fastest wave was found to correlate with the orientation of the elongated wound ($P<0.001$, $R^2 = 74\%$). Similarly, the orientation of  fastest wave was the same for both \emph{in vivo} and excised \emph{isolated}  samples, indicating that the STLs have a structural basis. Resulting wounds expanded by an average area of 9\% ($+16\%$ along STL and $-10\%$ across) while excised skin shrunk by an average area of 33\% ($23\%$ along STL and $10\%$ across).
\\[4pt] 
\emph{Conclusion}\\
Elastic surface wave propagation has been validated experimentally as a robust method for determining the orientation of STLs non-destructi\-ve\-ly and non-invasively. This study has implications for the identification of STLs and for the prediction of skin tension levels, both important factors in both human and veterinary reconstructive surgery.
\\[4pt]

\end{abstract}
\pagebreak


\section{Introduction}
\label{intro}


Early work on the mechanical behaviour of skin focused on the \emph{destructive} testing of skin samples, which had to be harvested from human cadavers \cite{Jansen58a, Ridge66a, Daly82, Dunn83, Stark77, Vogel87, Reihsner96} or from animal models \cite{Lanir79, Haut89, Eshel01, Shergold04, Liu08}. However, resecting an area of skin from a body dehydrates the sample and releases most of its residual stress, which is likely to alter its behaviour significantly.  Ideally, biological materials should be characterised in their physiological environment, where the biological and mechanical interactions with surrounding tissues and fluids are included. However, the possibilities for \emph{in vivo} testing are limited by technical feasibility oand by ethical considerations. When attempts are made at characterising human skin \emph{in vivo} through mechanical means (stretching \cite{Khatyr04, Flynn11a}, suction \cite{Diridollou00, Hendriks06, Delalleau08b}, torsion \cite{Agache80, Batisse08}, indentation \cite{Pailler08}, etc.), the skin is distorted, rotated and/or stretched, which briefly deforms the collagen fibre network and thus, significantly affects the accuracy of the experimental measurements.

In this study, we perform \emph{in vivo} testing by relying on the performance of the Re\-vis\-co\-me\-ter\textsuperscript{\textregistered}, a commercial device designed to test the human skin \emph{in vivo}, non-destructively and non-invasively \cite{Rev}. To date, the device has typically been used in the comparative assessment of topical cosmetic products \cite{Paye07} but it has also been used in several clinical studies for measurements on normal skin, diseased skin and scars \cite{Quatresooz08, Ruvolo07,Verhaegen10}. These latter studies have suggested that the Reviscometer\textsuperscript{\textregistered} can identify the direction of Skin Tension Lines (STLs) or Langer Lines. Since their discovery by Karl Langer in the 19th Century, these lines have been used by surgeons to decide the location and orientation of an incision. Langer observed that originally circular wounds became elliptical due to inherent tension in the skin. By creating multiple circular wounds on a cadaver and drawing lines along the major axis of the resulting ellipses, the Langer Lines were born. The Lines can be considered to be the lines of maximum \emph{in vivo} tension in the skin, which form a complex map over the body. To reduce the likelihood of a wound tension, dehisence (wound rupture) and a subsequent unsightly scar, incisions should be parallel to Langer lines, which lie along the path of maximum skin tension \cite{Borges84}. However, the orientation of Langer Lines are patient specific and depend on location, age, health, body mass index, ethnicity and hydration. No accepted method for determining STLs exists \emph{in vivo}, and most surgeons rely on generic maps or the pinch test to determine their orientation approximately. 

The Reviscometer\textsuperscript{\textregistered} operates by recording the time it takes for the wavefront of an elastic disturbance to travel over 2mm of skin. By rotating the direction of its probe and spanning 360$^{\circ}$, it can reveal directions of fast and slow wave propagation. The direction of the fastest wavefront has previously been shown to correlate with the direction of the STL \cite{Quatresooz08, Liang10}. As the preferred orientation of the collagen fibres corresponds to the STLs \cite{NiAnnaidh12b}, the speed of propagation of elastic disturbances on the skin can also, in principle, determine the orientation of collagen fibres. 

Canine skin has the same fundamental composition as human skin but has a different structure \cite{Scott01}. 
Canine skin is thicker than that of human skin: the thickness of canine epidermis is between 0.1 to 0.5 mm and the thickness of canine dermis is 0.8 to 5mm, whereas the thickness of human epidermis is between 0.05 to 0.1 mm and the thickness of human dermis is between 0.5 to 5mm \cite{Pavletic93, Scott01, MacGrath10}. 
Deroy \cite{Deroy14} and N\'i Annaidh et al. \cite{NiAnnaidh12a} determined through uniaxial extension tests that canine skin and human skin exhibit mechanical non-linear behaviour and a strain stiffening effect as the collagen fibres progressively align with the direction of applied tension.
N\'i Annaidh et al. \cite{NiAnnaidh12a} reported elastic moduli for human skin as $51\pm31$ MPa (transverse) and $93\pm28$ MPa (longitudinal) and Deroy \cite{Deroy14} reported elastic moduli for canine skin as $214$ MPa (transverse) and $432$ MPa (longitudinal). 
While the stiffness of the skin differs between humans and canines, it is clear that their fundamental structure and mechanical behaviour are similar. 
The mechanical properties of canine skin are important for veterinary dermatology and surgery and for its use as an animal model for human skin disease. 
Similar to the STLs described for humans, tension lines in canine skin have also been described \cite{Irwin66}. 
As in humans, STLs in dogs serve as a general guideline for incision creation/closure and flaps/grafts \cite{Pavletic10}. 
Furthermore, healing of sutured wounds in dogs can be influenced by tension, pressure, motion and patient health. 
In veterinary reconstructive surgery, when a segment of skin is completely removed from the body and transferred to a recipient site, it undergoes significant deformation. 
For the reasons outlined here, canine skin has been chosen as a suitable model for human skin, not only because the mechanical properties are comparable, but also because the clinical needs in both veterinary and human surgery are similar. 

The objectives of the present study are therefore to (a) validate the Reviscometer\textsuperscript{\textregistered} using two independent methods for determining the STLs and (b) to improve the understanding of the \emph{in vivo} mechanical behaviour of canine skin for surgical planning. We hypothesized and confirmed that the preferred orientation of collagen fibres (and therefore the greatest stiffness) is aligned with the STLs. Furthermore, we hypothesized and confirmed that the STLs determined with the Reviscometer\textsuperscript{\textregistered} are the same as those determined with Langer's original test. Using this technique, it is possible to determine, \emph{in vivo}, the orientation of patient-specific STLs, non-invasively and in real time. 
This study has clear implications for the identification of STLs and the prediction of skin tension levels, both important factors in preoperative planning for reconstructive surgeries \cite{Tepole15}.


\section{Materials and Methods}
\label{Methods}



\subsection{Method of operation}
\label{operation}

\begin{figure}[ht!]
\centering
\subfigure[]{\includegraphics[width=0.45\textwidth]{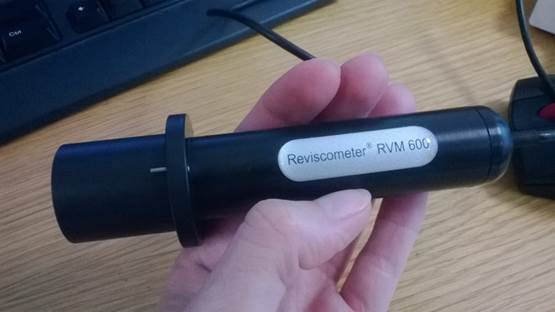}}
\subfigure[]{\includegraphics[trim = 45mm 0mm 50mm 0mm, clip,width=0.45\textwidth]{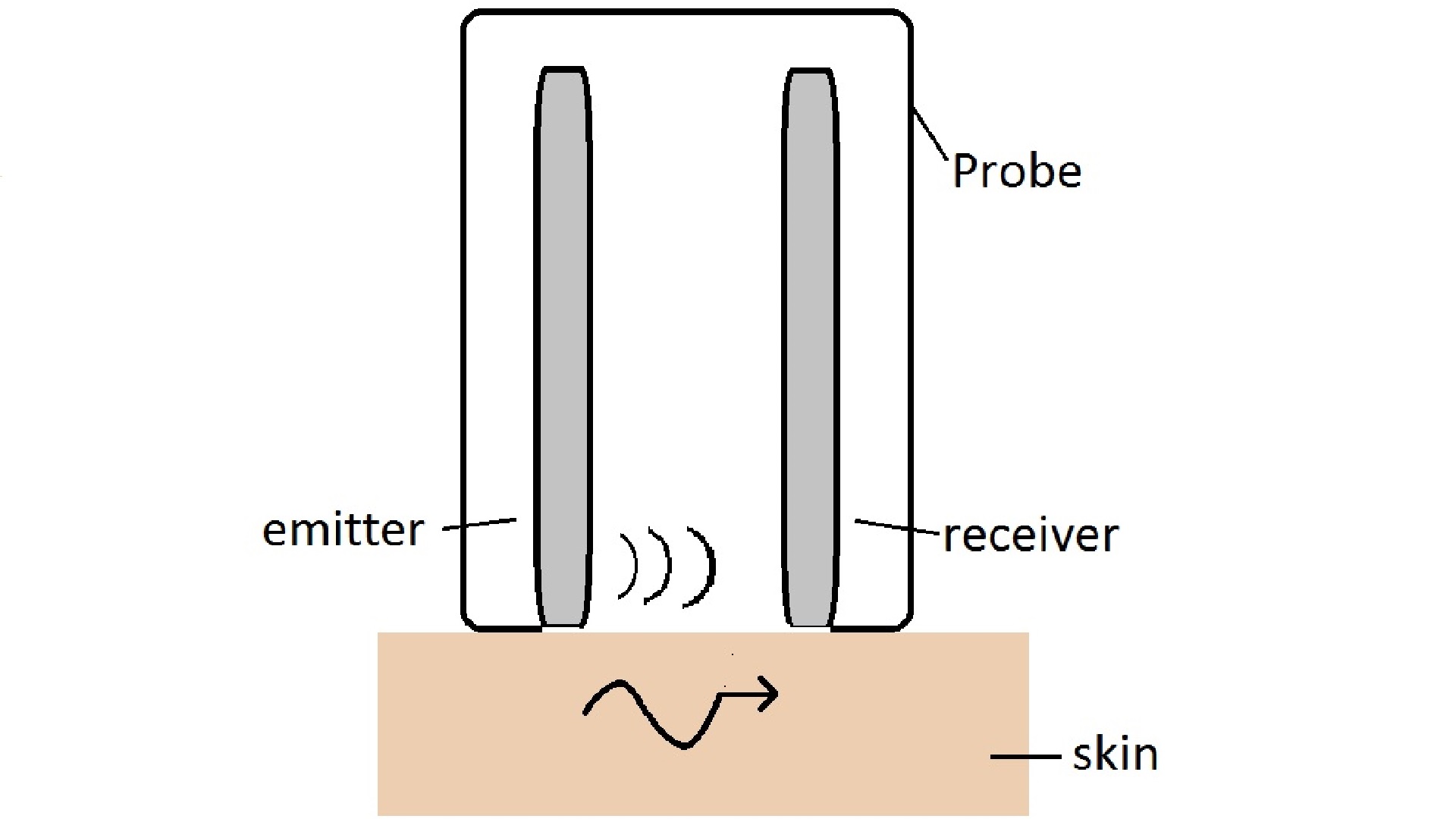}}
\caption{
\footnotesize{(a) Reviscometer device\textsuperscript{\textregistered}. (b) Mechanism of operation.}}
\label{reviscometer}
\end{figure}

The Reviscometer\textsuperscript{\textregistered} device (Courage \& Khazaka Electronic GmbH, Köln, Germany), shown in Fig \ref{reviscometer}(a), is composed of a probe with two needle sensors placed 2 mm apart, put into contact with skin \emph{in vivo}. 
The first needle emits an acoustic shock-wave by impacting the skin with a force of 1N and the second needle acts as a receiver as shown in Fig \ref{reviscometer}(b).
The device measures the time taken to travel the 2 mm distance, known as the ``Resonance Running Time'' (RRT), and gives direct access to the speed of propagation of a surface wave in the skin. All measurements are recorded using the integrated MAP5 software.
While the device only provides measurements along one direction, the probe is rotated in $10^{\circ}$ increments, so that thirty six different measurements are taken over the full  $360^\circ$ rotation, giving the variation of the wave speed with the angle, and demonstrating the level of anisotropy for the skin. 

\subsection{Specimen preparation}
\label{Animal methods}


Animal testing was performed at University College Dublin (UCD) on the cadavers of two healthy young adult dogs that were euthanized for reasons unrelated to this study. 
The cadaver dogs were frozen at $-18^\circ$C  and allowed to thaw at $20^\circ$C  for 48 hours. 
Physical examinations were performed to ensure that the dogs had no evidence of skin lesions. 
All results were obtained from normal, healthy skin free of dermatological irregularities in the area investigated. 
Skin from the back of the dog was prepared in a similar manner as described by Bismuth et al. \cite{Bismuth14}. 
The cadavers were shaved in the region of interest with clippers (Aesculap\textsuperscript{\textregistered}) and skin was degreased using diethyl ether and absorbent paper. 
Cadaver 1 was a five year old female, mixed-breed, medium sized dog (11 kg) and Cadaver 2 was a four year old male greyhound (28 kg). 
During testing, the cadavers were placed in lateral recumbency in a physiologically normal position. 
The choice of test site was chosen so as to avoid pressure from underlying bony prominences and to ensure subsequent incisions did not interfere with shape of the wound. 
This was achieved by ensuring the distance between incisions was at least 100\% of the size of the incision. The test sites were then marked using surgical secureline skin markers (Cardinal Health\textsuperscript{\textregistered}).


\subsection{Skin Tension Lines identification via surface wave propagation}
\label{testing}


All measurements were performed at room temperature ($20^\circ$C) with the Reviscometer\textsuperscript{\textregistered}. 
Thirty six measurements were taken from $0$ to $360^\circ$ and were repeated three times at each test site by the same individual. 
Following testing, the collected data was analysed and the orientation of the fastest travelling acoustic wave was determined and indicated on the cadaver. 
A circular excision was then made at each of the test locations. 
It was expected that the wound would become ellipsoidal, as originally described by Langer \cite{Langer78}, indicating the direction of STLs. 
The orientation of ellipses was determined via image analysis (described further in Section \ref{imaging}), which were then compared to the direction predicted by RRT measurements based on the wave speed. 
 
\begin{figure}[ht!]
\centering
\subfigure[]{\includegraphics[trim={0.5cm 0cm 1cm .3cm},clip,width=0.45\textwidth]{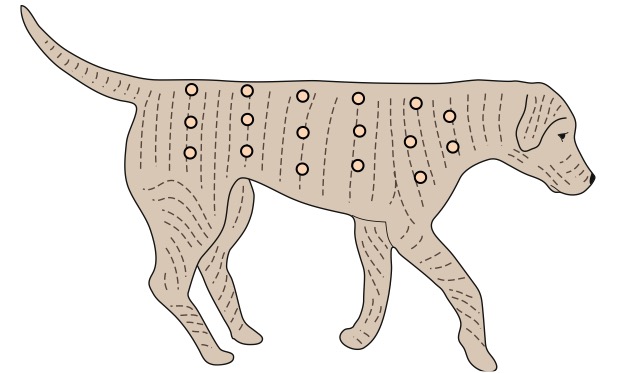}}
\subfigure[]{\includegraphics[trim={0.5cm 0.5cm .5cm 0cm},clip,width=0.25\textwidth, angle=90]{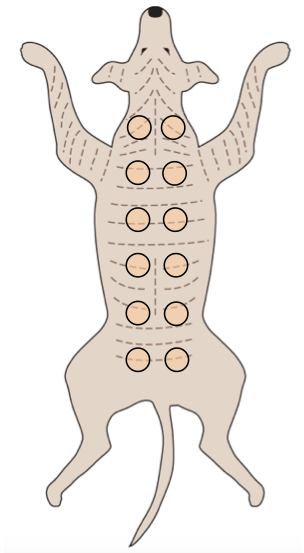}}
\caption{
\footnotesize{Test location on (a) Cadaver 1 and (b) Cadaver 2 (amended from \cite{Tobias12})}}
\label{biopsy}
\end{figure}

In the case of Cadaver 1, seventeen test locations were marked on the skin between the trunk and the neck, as indicated in Fig. \ref{biopsy}(a). 
A circular biopsy punch (Medline Industries\textsuperscript{\textregistered}) of 8mm diameter was then used to make the incisions similar to the method described by Oiki \cite{Oiki03}.

In the case of Cadaver 2, twelve test locations were selected on the trunk and were marked  as indicated in Fig.\ref{biopsy}(b).
In this case, a scalpel and a plastic template were used to make the larger circular incisions of 6cm in diameter as in \cite{Upchurch14}. 
The excised (full thickness) skin sample, including the epidermis, dermis and hypodermis, was placed on a wet impermeable plexiglass surface (to reduce surface friction). 
RRT measurements were then performed on the excised skin sample and RRT measurements between the \emph{in situ} sample and the corresponding isolated excised skin sample were compared. 
Five minutes after skin excision, the expansion / contraction of both the excised sample and the resulting wound were photographed against underlying graph paper.


\subsection{Image analysis of wound morphology}
\label{imaging}

\begin{figure}[ht!]
\centering
\subfigure[]{\includegraphics[trim={3cm 2.5cm 3cm 1.5cm},clip, width=0.3\textwidth]{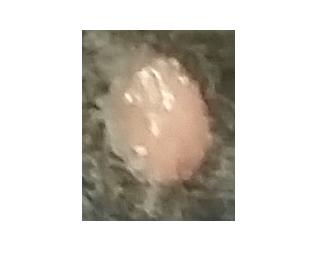}}
\subfigure[]{\includegraphics[trim={3cm 2.5cm 3cm 1.5cm},clip,width=0.3\textwidth]{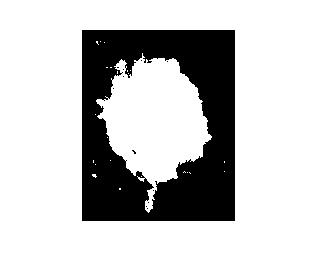}}
\subfigure[]{\includegraphics[trim={3cm 2.5cm 3cm 1.5cm},clip,width=0.3\textwidth]{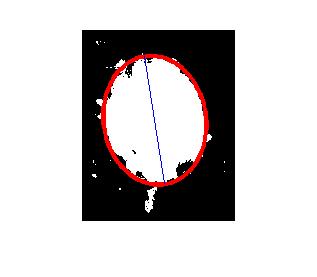}}
\caption{
\footnotesize{(a) Original image of \emph{in situ} wound (b) Images of each wound were binarised based on an automated threshold level producing an image containing white pixels for the wound and black pixels for all other areas; (c) The $regionprops$ function fit an ellipse (red) to each wound by matching the second order moments of
that component to an equivalent ellipse. The $regionprops$ function also output a set of properties for each ellipse including the orientation of the major axis (blue) and the geometry of the ellipse.} }
\label{image}
\end{figure}

Elliptical distortion of wounds and excisions was observed in most cases. The orientation and level of expansion/contraction of wounds with respect to the STLs were then calculated by fitting a best fit ellipse in a customised MATLAB routine using the Image Processing Toolbox (see Fig. \ref{image}).
The advantage of fitting an ellipse about each component is that the orientation is measured in a systematic and repeatable manner
which can account for the non-uniformities of the shape of each component.

\subsection{\emph{In vivo} human testing}
\label{human methods}


Following verification that the described technique can determine the orientation of tension lines within the skin, measurements were taken from a healthy 30 year old female volunteer. Ethical approval was obtained from the Human Research Ethics Committee of UCD and the study complied fully with the Helsinki Declaration. As before, RRT measurements were taken in $10^\circ$ increments about $360^\circ$ on the volar forearm. The mean RRT and anisotropic ratio (ratio of maximum RRT to minimum RRT) of human and canine skin were then compared.


\section{Results}
\subsection {Non-invasive identification of Skin Tension Lines}
\label{LL identification}

\begin{figure}[ht!]
\centering
\subfigure[]{\includegraphics[trim={4cm 2cm 3.8cm 3.5cm},clip, width=0.45\textwidth]{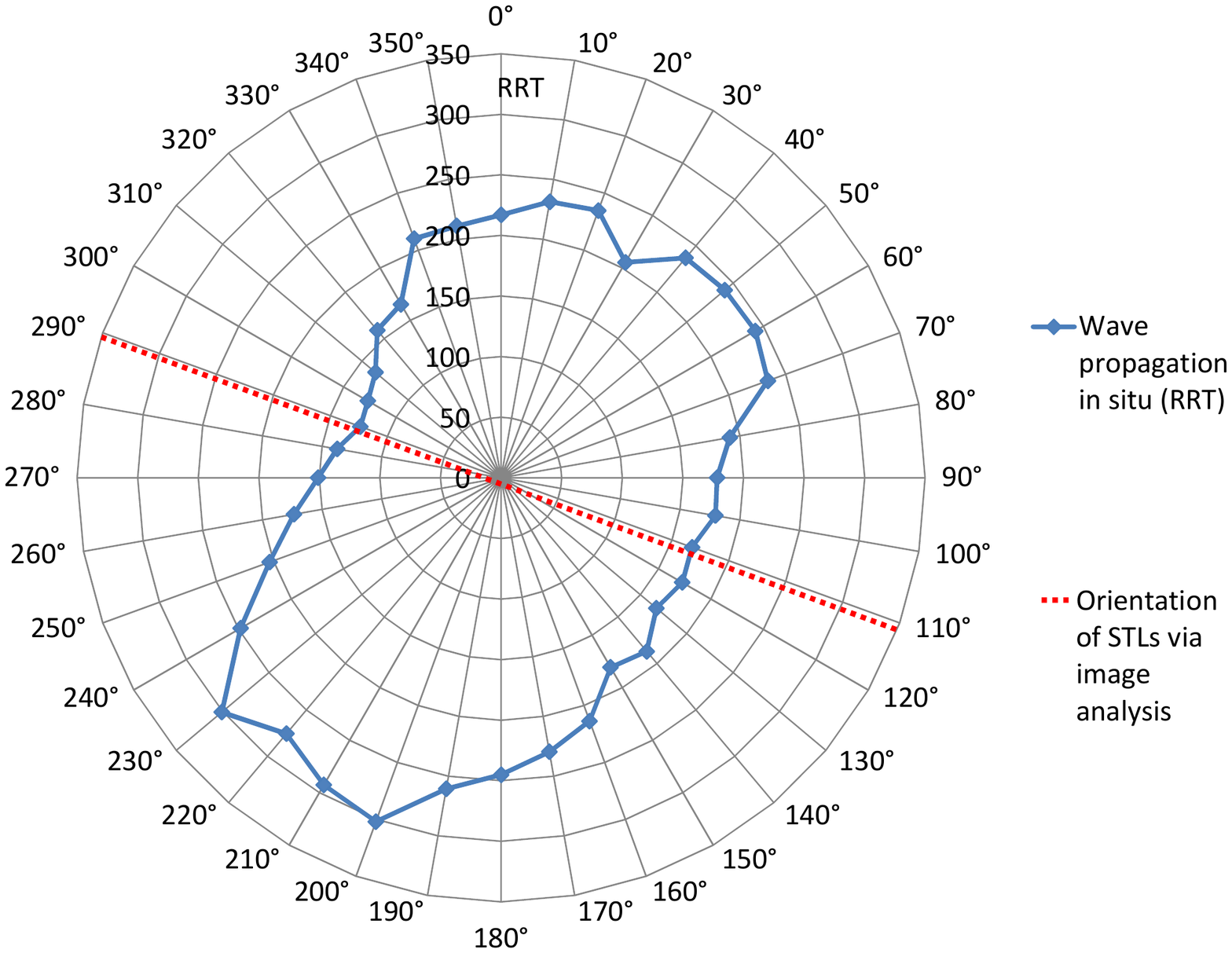} }
\subfigure[]{\includegraphics[trim={6.5cm 7cm 5cm 7cm},clip,width=0.45\textwidth]{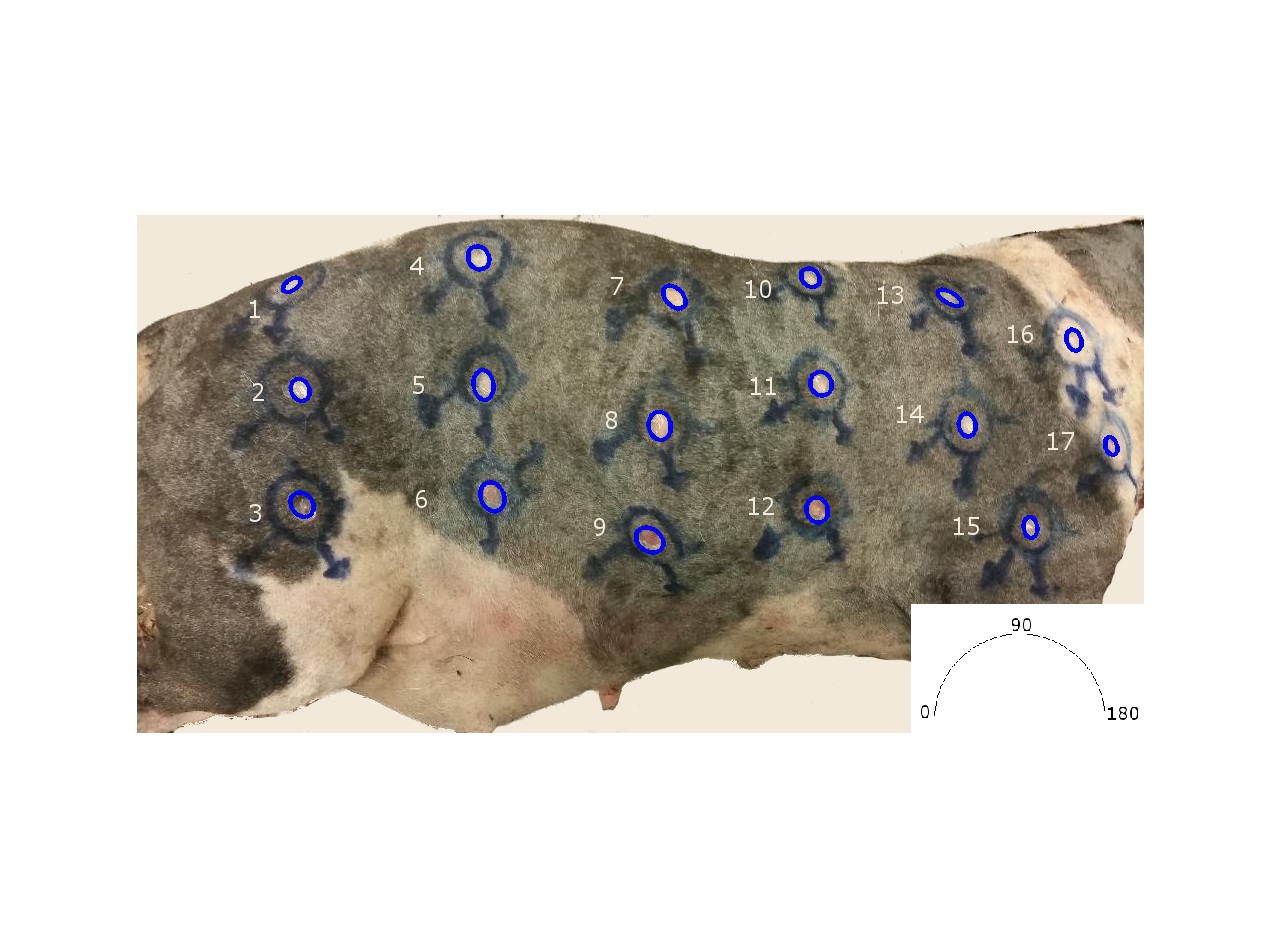}}
\caption{
\footnotesize{
(a) Mean Resonance Running Time (RRT) plotted against the orientation of the probe (blue) for a representative example. The orientation of STLs determined invasively is indicated by the dotted line. The lowest RRT (i.e. fastest wave speed) corresponds to the orientation of STLs. (b) Best fit ellipses determined through image analysis (blue) of each wound. }\label{LLID}}
\end{figure}

Dog cadaver experiments revealed a number of interesting facts about the use of the Reviscometer\textsuperscript{\textregistered}. 
Firstly, it can be seen that the wave speed varies with respect to the orientation of the probe (as shown in Fig. \ref{LLID}(a)). 
We propose that the surface wave travels fastest along the direction of highest tension and slowest in the direction of lowest skin tension. 
By identifying the orientation of the fastest travelling wave, we can therefore identify the direction of STLs. 
The non-invasive prediction of STLs using the Reviscometer\textsuperscript{\textregistered} was then compared to the invasive prediction using image analysis of the wounds as shown in \ref{LLID}(b). Details of measurements are given in Table \ref{table:orientation}. 
A significant correlation ($P<0.001$, $R^2 = 74\%$) exists between the two independent measurement techniques; 
however, a Bland-Altman analysis indicated that an average bias of $13^\circ$ exists between the methods with a 95\% confidence interval of $+52^\circ$ and $-22^\circ$. 
Finally, the orientation of STLs in dogs was compared to historical generic maps from \cite{Irwin66} and were found to be in broad agreement. 
These results indicate that the patient specific orientation of STLs can be determined non-invasively and in real time using the Reviscometer\textsuperscript{\textregistered}, albeit with an uncertainty which will be discussed further in Section \ref{discussion}.

\begin{table}[h!]
  \begin{center}
    \caption{
        Average orientation measurements with Reviscometer\textsuperscript{\textregistered} and image analysis. \label{table:orientation}}
        \vspace{18pt}
    \begin{tabular}{ccc}
      \toprule
      Sample		& Image Analysis	&Reviscometer \\
      \midrule
    
		1&	146$^{\circ}$&	150$^{\circ}$\\
2&	64$^{\circ}$&80$^{\circ}$\\
3&	51$^{\circ}$&	30$^{\circ}$\\
4&	61$^{\circ}$&	40$^{\circ}$\\
5&	82$^{\circ}$&	170$^{\circ}$\\
6&	67$^{\circ}$&	30$^{\circ}$\\
7&	46$^{\circ}$&	30$^{\circ}$\\
8&	81$^{\circ}$&	0$^{\circ}$\\
9&	34$^{\circ}$&	10$^{\circ}$\\
10&	42$^{\circ}$&	20$^{\circ}$\\
11&	65$^{\circ}$&	40$^{\circ}$\\
12&	71$^{\circ}$&	20$^{\circ}$\\
13&	29$^{\circ}$&	30$^{\circ}$\\
14&	81$^{\circ}$&	90$^{\circ}$\\
15&	84$^{\circ}$&	70$^{\circ}$\\
16&	71$^{\circ}$&	90$^{\circ}$\\
17&	71$^{\circ}$&	50$^{\circ}$\\
      \bottomrule
    \end{tabular}
  \end{center}
\end{table}

\subsection {Alignment of collagen fibres with Skin Tension Lines}
\label{fibre alignment}

\begin{figure}[ht!]
\centering
\includegraphics[trim={4cm 2cm 3.8cm 3.5cm}, clip, width=0.6\textwidth]{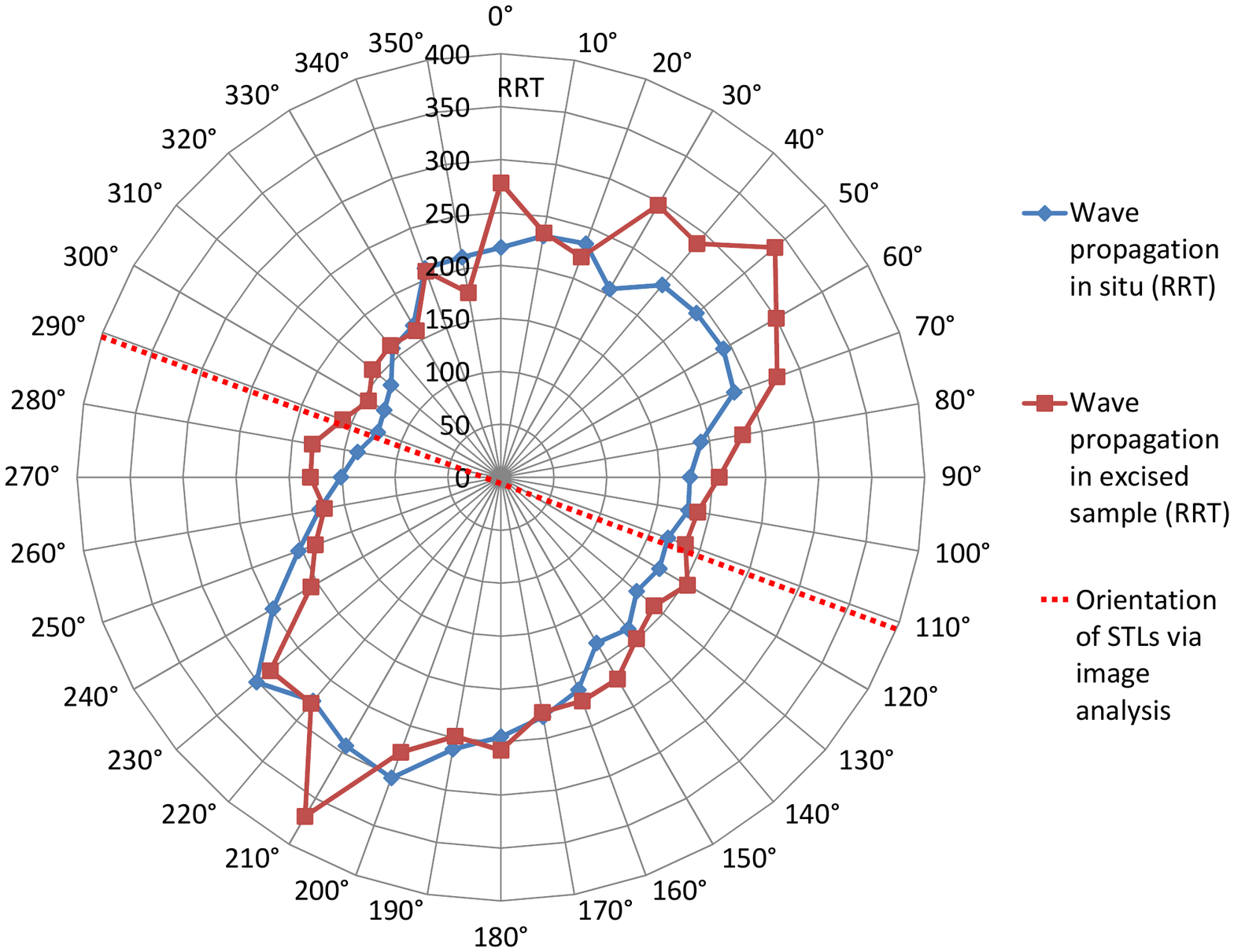} 
\caption{
\footnotesize{
Wave speed of an \emph{in situ} and \emph{isolated} skin sample compared with the orientation of STLs determined through image analysis}}
\label{isolated}
\end{figure}

A paired t-test revealed that the identified orientation of STLs of \emph{in situ} samples and their corresponding $isolated$ excised skin samples were not statistically different (P=0.21). 
These tests also revealed that the magnitudes of anisotropy were not statistically different (P=0.21) and are comparable as shown in Fig.\ref{isolated}. 
Therefore, since the same anisotropy is present even once skin tension is released, these results indicate that the observed anisotropy is, at least in part, due to the structure of the skin itself, namely, the alignment of collagen fibres in the dermis. 
This also supports the hypothesis that the STLs are aligned with the predominant orientation of collagen fibres.

\subsection {Contraction and expansion of wounds}
\label{contraction}


As shown in Table \ref{table:expansion}, the average area expansion of the circular wounds was 9\%, corresponding to a line expansion of 16\% parallel to the STLs (by image analysis), but a line contraction of 10\% perpendicular to the STLs. Similarly, the area of the excised sample shrank by 33\% overall, shrinking 23\% parallel to the STLs and shrinking 10\% perpendicular to the STLs. Fig. \ref{contraction} provides a summary of the levels of expansion and contraction of both the excised skin sample and the resulting wound. 

\begin{table}[h!]
  \begin{center}
    \caption{
    Average expansion (+) and contraction (-) of wounds and \emph{isolated} samples along and across the Skin Tension Lines (STL). \label{table:expansion}}
  \vspace{18pt}
    \begin{tabular}{cccc}
      \toprule
      Sample		& Average	&Along STL	& Across STL \\
      \midrule
     Wound	& +9\%	& +16\%		&-10\% \\
		Excised	& -33\%	& -23\%		&-10\% \\
      \bottomrule
    \end{tabular}
  \end{center}
\end{table}

\begin{figure}[ht!]
\centering
\includegraphics[trim={3.1cm 0cm 1cm 0.2cm},clip, width=0.6\textwidth]{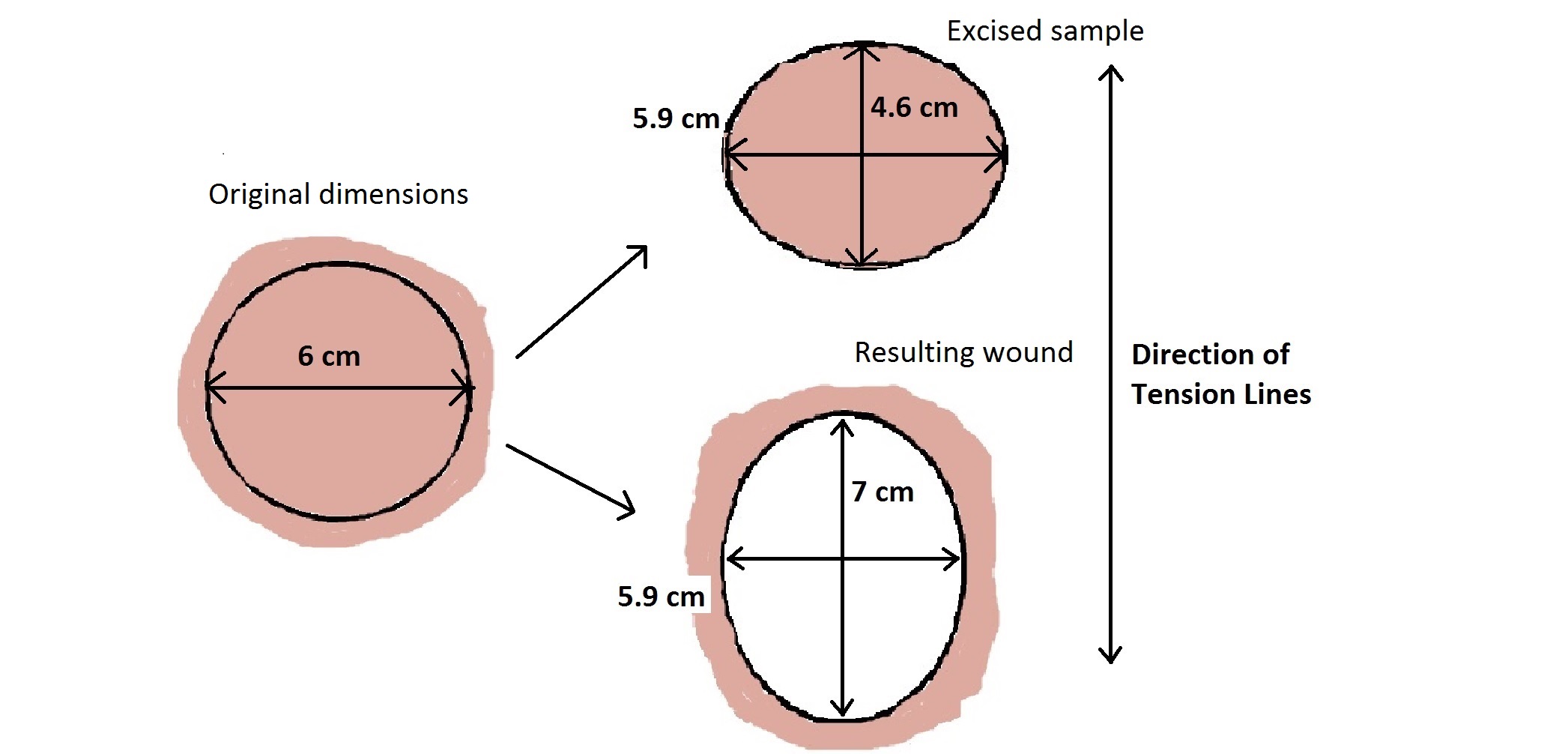} 
\caption{\footnotesize{
Illustration of average contraction and expansion levels for 1) an \emph{isolated} excised sample and 2) and originally circular wound}}
\label{contraction}
\end{figure}


\subsection {\emph{In vivo} human testing}
\label{Human Results}

\begin{figure}[ht!]
\centering
\includegraphics[trim={3.1cm 2.3cm 4cm 2.2cm},clip,width=0.6\textwidth]{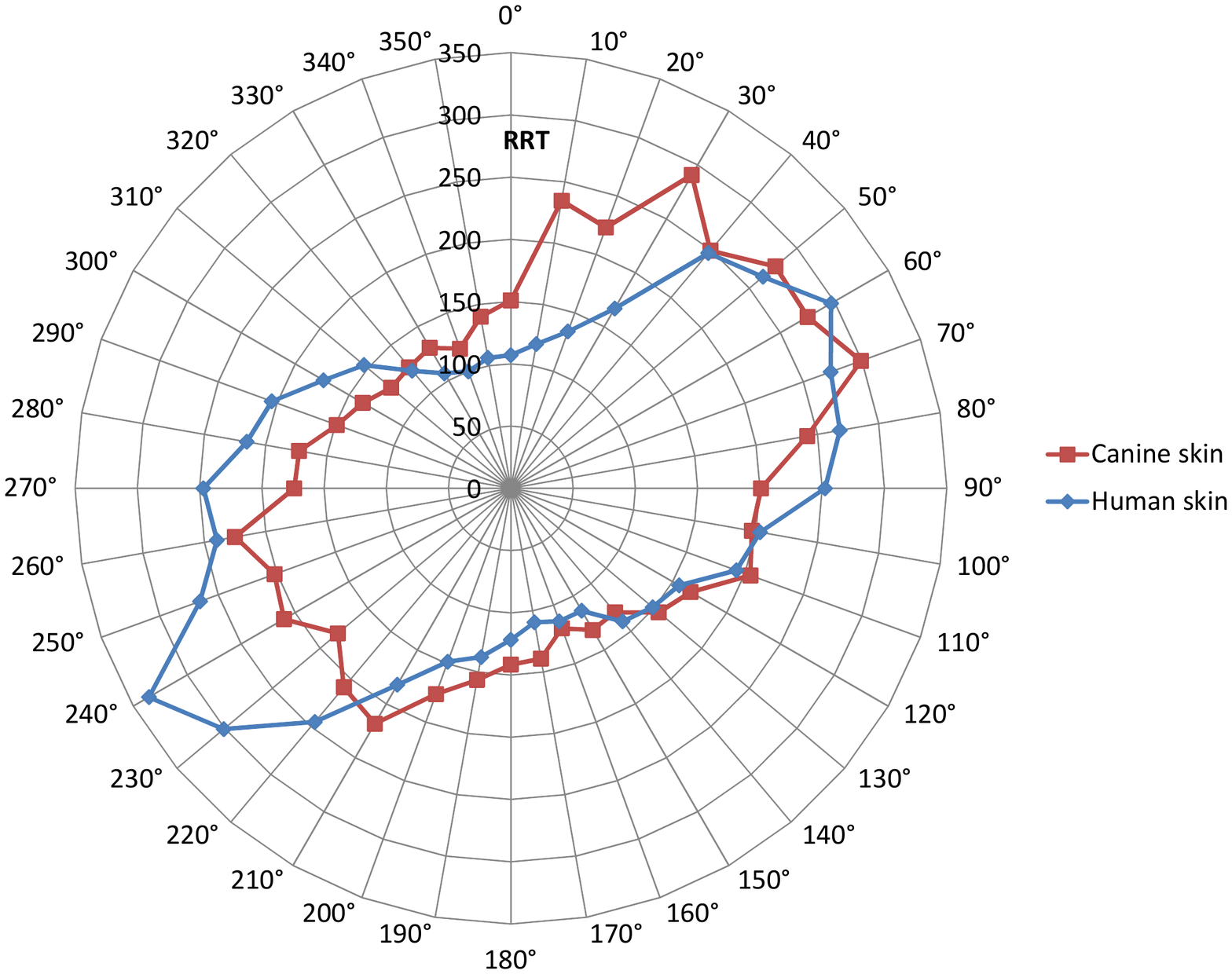}
\caption{Mean Resonance Running Time (RRT) plotted against the orientation of the probe for both human and canine skin.}
\label{fig:aniso}
\end{figure}

Preliminary \emph{in vivo} testing shown in Fig. \ref{fig:aniso} revealed that the Reviscometer\textsuperscript{\textregistered} provides similar results for both human and canine skin. It was found that the average magnitude of RRT differed by only 15\% and the level of anisotropy differed by 26\%, indicating that 1) canine skin is a suitable animal model for validating the device and 2) that the device is suitable for use in veterinary medicine.

\section {Discussion}
\label{discussion}


In this study, we provide and validate new experimental data of \emph{in vivo} canine skin focusing in particular on the anisotropic properties. 
We quantify and compare the degree of anisotropy of the skin with respect to STL orientation using two independent experimental techniques: a non-invasive method based on acoustic wave propagation and an invasive method adapted from Langer's original experiments \cite{Langer78}. 
By validating the non-invasive technique (Reviscometer\textsuperscript{\textregistered}) experimentally, we have demonstrated that this method can be used to identify STL patterns all over the body, thereby providing a simple and effective means to gather basic information for preoperative surgical planning. 
Similar claims have been previously made by various authors \cite{Quatresooz08, Ruvolo07}, with Ruvolo et al.  \cite{Ruvolo07} in particular, finding that the RRT value is inversely related to the skin's elasticity i.e. the higher the RRT, the lower the stiffness of the skin. 
However we are the first group to validate the Reviscometer\textsuperscript{\textregistered} method using the original experimental technique outlined by Langer \cite{Langer78}.

Similar to the generic STL maps developed for human plastic surgery \cite{Ridge66a, Langer78, Borges84}, tension lines in the dog have been previously described \cite{Irwin66, Oiki03}. 
These results, however, are not generalisable, as the elasticity and abundance of skin varies markedly between breeds and individuals. They can also vary on the basis of body region, age, body conformation, and pathologic conditions involving the cutaneous tissues \cite{Hedlund06, Swaim97, Pavletic10}. 
For these reasons, it is clear that patient-specific identification of STLs is required. 
However, no methods have been described for a non-invasive evaluation of skin tension lines, save the pinch test. 
The pinch test is performed by gently pinching the skin in the area of interest to form furrows. 
The direction of these furrows form the Relaxed Skin Tension Lines (RSTLs) \cite{Borges84}. 
This technique, however, can be subjective and relies on the expertise of the surgeon \cite{Seo13}. 
While there is ongoing debate as to which set of tension lines (RSTL, Langer, Kraissl etc.) is the most useful, it is clear that skin anisotropy and skin tension are important issues for surgeons \cite{Carmichael14}. 
While here we have focused our attention on lines of maximum skin tension, the method described can similarly determine lines of minimum skin tension which are orientated perpendicular to lines of maximum tension.

In addition to planning incisions, the Reviscometer\textsuperscript{\textregistered} may also prove useful for applications in skin grafts. Here, we have shown that the direction of maximum shrinkage of an excised piece of skin is aligned with the direction of minimum RRT. While further research is required to support the claim, it is proposed here that the Reviscometer\textsuperscript{\textregistered} can also be used to match the fibre orientation and relative levels of \emph{in vivo} skin tension of harvested skin with the target location. 

While the present study has important implications in surgery there are a number of limitations. Tests were performed on frozen canine cadavers of different breeds: the mechanical behaviour of skin is likely affected by the freezing/thawing process inducing changes in hydration levels and damage to the extracellular matrix \cite{Gianinni08}. Furthermore, the thawed tissue resulted in damp skin in places, which interferes with the Reviscometer\textsuperscript{\textregistered} readings and led to a number of outliers. 
While a statistically significant correlation ($P<0.001, R^2 = 74\%$) exists between the two independent measurement techniques, a Bland-Altman analysis revealed a 95\% confidence interval of $+52^\circ$ and $-22^\circ$. To interpret this high uncertainty, one must consider that the accuracy of both experimental techniques relies upon the anisotropic nature of skin to identify the STLs. Indeed, further analysis revealed that the accuracy of both techniques improved with increasing eccentricity (or levels of anisotropy). This should not be considered a limitation insofar as the direction of STLs is only important where a strong anisotropy exists.

In conclusion, the Reviscometer\textsuperscript{\textregistered} allows objective evaluation of STLs non invasively, offering the possibility of obtaining effective and reliable measurements for preoperative surgical planning.  Furthermore, the device could potentially be used in the training of surgeons. This research has  provided proof-of-concept using animal cadavers; However, ideally a clinical (human or animal) trial could be conducted to provide further confirmation of the technique in a surgical setting.



\bibliographystyle{vancouver}
\bibliography{references}


\end{document}